\documentclass[conference]{IEEEtran}
\usepackage{cite}
\usepackage{amsmath,amsthm,amssymb,amsfonts,bm}
\newtheorem{theorem}{Theorem}

\newtheorem{lemma}{Lemma}
\newtheorem{problem}{Problem}
\newtheorem{proposition}{Proposition}

\newtheorem{question}{Question}
\newtheorem{corollary}{Corollary}
\newtheorem{remark}{Remark}
\newtheorem{defn}{Definition}
\usepackage{algorithmic}
\usepackage{graphicx}
\usepackage{epstopdf}
\usepackage{textcomp}
\usepackage{xcolor}
\usepackage{url}
\usepackage{enumitem}
\usepackage{url}
\usepackage{bbm}
\setlist[itemize]{leftmargin=*}
\setlist[enumerate]{leftmargin=*}
\usepackage{multirow}
\newcommand{\tabincell}[2]{\begin{tabular}{@{}#1@{}}#2\end{tabular}} 
\usepackage{subfigure}
\usepackage[linesnumbered, ruled]{algorithm2e}
\def\BibTeX{{\rm B\kern-.05em{\sc i\kern-.025em b}\kern-.08em
    T\kern-.1667em\lower.7ex\hbox{E}\kern-.125emX}}
\begin{document}
\title{Incentive Mechanism Design for  Distributed  \\Coded Machine Learning}
\IEEEoverridecommandlockouts
\author{Ningning Ding, 
	Zhixuan Fang, Lingjie Duan, 
	and~Jianwei~Huang
	\thanks{This work is supported by the Shenzhen Institute of Artificial Intelligence and Robotics for Society, and the Presidential Fund from the Chinese University of Hong Kong, Shenzhen.
	}				
	\thanks{Ningning Ding is with the Department of Information Engineering, The Chinese University of Hong Kong, Hong Kong  (e-mail: dn018@ie.cuhk.edu.hk).}	
	\thanks{Zhixuan Fang is with  Institute for Interdisciplinary Information Sciences, Tsinghua University, Beijing, China, and Shanghai Qi Zhi Institute, Shanghai, China (e-mail: zfang@mail.tsinghua.edu.cn). }	
	\thanks{Lingjie Duan is with  the Engineering Systems and Design Pillar, Singapore University of Technology and Design, Singapore (e-mail: lingjie duan@sutd.edu.sg).}
	\thanks{Jianwei Huang is with the School of Science and Engineering, The Chinese University of Hong Kong, Shenzhen, and the Shenzhen Institute of Artificial Intelligence and Robotics for Society (corresponding author, e-mail: jianweihuang@cuhk.edu.cn).}	
	}

\maketitle
\begin{abstract}
	A distributed machine learning platform needs to recruit many heterogeneous worker nodes to finish computation simultaneously. As a result, the overall performance may be degraded due to straggling workers. By introducing redundancy into computation, coded machine learning can effectively improve the runtime performance by recovering the final computation result through the first $k$ (out of the total $n$) workers who finish computation. While existing studies focus on designing efficient coding schemes, the issue of designing proper incentives to encourage worker participation is still under-explored. This paper studies the platform's optimal incentive mechanism for motivating proper workers' participation in coded machine learning, despite the incomplete information about heterogeneous workers' computation performances and costs. A key contribution of this work is to summarize workers' multi-dimensional heterogeneity as a one-dimensional metric, which guides the platform's efficient selection of workers under incomplete information with a linear computation complexity. Moreover, we prove that the optimal recovery threshold $k$ is linearly proportional to the participator number $n$ if we use the widely adopted MDS (Maximum Distance Separable) codes for data encoding. We also show that the platform's increased cost due to incomplete information disappears when worker number is sufficiently large, but it does not monotonically decrease in worker number.
\end{abstract}


\section{Introduction}
\subsection{Background and Motivations}
Recent years have witnessed the rapid development of large-scale distributed machine learning, which well reconciles the massive computation tasks and the limited computation power of a single worker's  machine. 
Many  workers (e.g., using servers, laptops, or even smartphones) can locally conduct model training based on assigned training data, and they feed back the computation results to the platform  to complete a larger machine learning task. 
However, the  performance of large-scale distributed machine learning systems can be significantly affected by bottlenecks like straggling workers which finish computation much slower than other workers \cite{dean2013tail}. 

A promising solution to  effectively alleviate straggler bottlenecks is to use coding techniques to create proper computation redundancy \cite{lee2017speeding}. 
The key idea of coded machine learning is to properly encode the data to be used for the workers'  computation subtasks, such that  the platform only needs the computation results from a subset $k$ workers among a total of $n$ workers who are assigned the subtasks, in order to complete the overall computation task.   Hence, the fastest subset of workers will determine   the overall runtime of the distributed machine learning.  The minimum number of workers that the platform needs to wait for (i.e., $k$) is called  \emph{recovery threshold}, which   depends on the code design and subtask assignment.  
 
 \begin{figure}[tbp]
 	\centering
 	\subfigure[Encoding and  assignment]{	
 		\begin{minipage}[t]{0.42\linewidth}
 			\centering
 			\includegraphics[width=1.2 in]{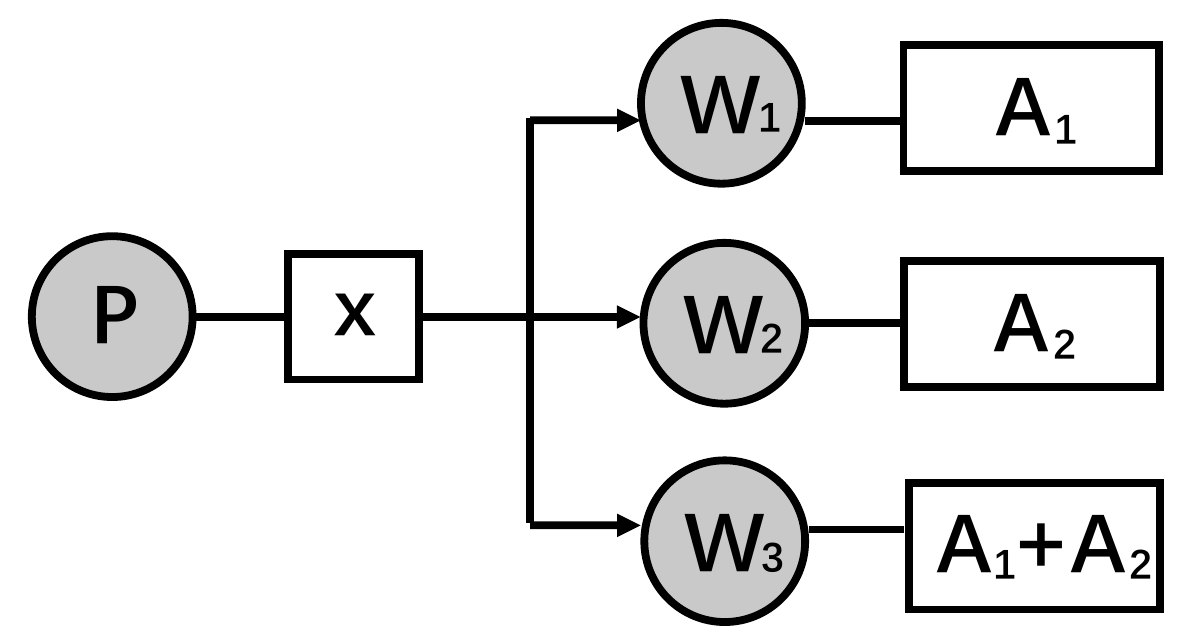}
 			\label{fig0}
 	\end{minipage}}
 	\subfigure[Computing and decoding]{
 		\begin{minipage}[t]{0.54\linewidth}
 			\centering
 			\includegraphics[width=1.8 in]{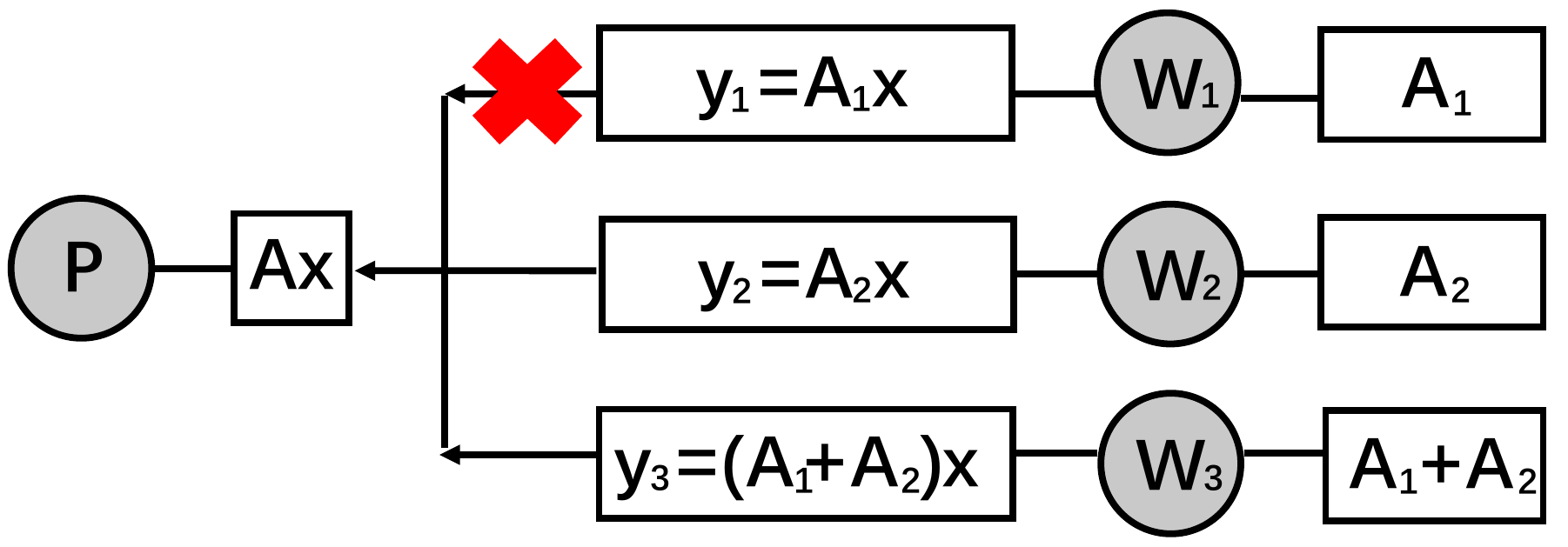}
 			\label{fig1}
 	\end{minipage}}
 	\caption{Illustration of  coded machine learning using $(n=3, k=2)$-MDS codes, which can recover the final computation result from only two (out of three)  workers' computation results.}
 	\label{cc}
 	\vspace{-6mm}
 \end{figure}

Fig.~\ref{cc} shows an illustrative  example  of coded machine learning, where we consider  a   system with one platform and  $n=3$  workers. The platform wants to compute a  matrix-vector multiplication $\mathbf{y}=\mathbf{A} \mathbf{x}$ using $(n=3, k=2)$-MDS  codes. 
\begin{itemize}
	 \item \emph{Encoding and assignment:} First, the platform splits matrix $\mathbf{A}\in \mathbb{R}^{r \times s}$ equally into two   matrices  $\mathbf{A}_1 \in \mathbb{R}^{\frac{r}{2} \times s}$ and  $\mathbf{A}_2\in \mathbb{R}^{\frac{r}{2} \times s}$. Then, the platform assigns matrix $\mathbf{A}_1$   to worker 1,   $\mathbf{A}_2$  to worker 2, and    $\mathbf{A}_1+\mathbf{A}_2$   to worker 3.  The platform  also broadcasts  the input vector $\mathbf{x}$ to  the three workers.
	
	\item \emph{Computing and decoding:}    The three workers will   compute subtasks  $\mathbf{y}_{1}=\mathbf{A}_{1} \mathbf{x}$, $\mathbf{y}_{2}=\mathbf{A}_{2} \mathbf{x}$, and $\mathbf{y}_{3}=(\mathbf{A}_{1}+\mathbf{A}_{2}) \mathbf{x}$ in parallel, respectively,   and return the  results to the platform  as soon as each worker finishes his own computation. The platform is able to obtain the intended outcome $\mathbf{A} \mathbf{x}$ as long as he receives any $k=2$ workers' results. For example, if worker 1 is the straggling worker which finishes the computation subtask the last,    the platform can simply obtain the intended outcome from workers 2 and 3, i.e.,  $\mathbf{y}=\left[\mathbf{y}_{3}-\mathbf{y}_{2} ; \mathbf{y}_{2}\right]=\left[\mathbf{A}_{1} \mathbf{x}; \mathbf{A}_{2} \mathbf{x}\right]=\mathbf{A} \mathbf{x}$. 
\end{itemize} 


However, given significant runtime saving using codes,   it   may not be realistic that workers are willing to participate in   coded machine learning without proper incentives, as workers are selfish and incur costs during the computation   \cite{duan2012incentive}. For example, a worker needs to postpone  his own background task for running the platform's assigned task, which translates to time costs locally.   

There are  a few challenges   in incentivizing workers' participation. 
First, it is challenging for the platform to  incentivize only the desirable workers to participate in coded machine learning, as the platform may not know workers' computation costs and computation time.  Appropriate  incentives    should be sufficient to cover the   costs of desirable workers and  be based on workers' computation time performance.  The computation costs can be workers' private valuation information, which may not be easily accessed by the platform due to workers' privacy concerns.  The worker's computation time  depends on many factors such as computation power as well as the unpredictable and unreliable computation infrastructure, hence it may not be even accurately known by the worker himself, not to mention the platform \cite{yang2019timely}. The joint consideration of private information and stochastic information is a key feature of our model, which differs from many incentive  mechanism design problems where there is only private and accurate information (to the workers)  (e.g., \cite{ningwiopt,ma2018incentivizing}).
This will lead to the first key question of this paper: 
\begin{question}
	How to incentivize heterogeneous workers with  private cost information and stochastic computation time to participate in coded machine learning?
\end{question}


Second, it is under-explored  of the impact of workers' private information on the platform's incentive mechanism design and cost. 
In the presence of workers' private information,   the platform may obtain more information through  market research. Different levels of information asymmetry require the platform to design different optimal   incentive mechanisms  and thus make the platform obtain different costs. It is under-explored how much cost savings the platform can obtain if he has  more information about workers. 
This motivates us to ask the second key question in this paper:  
\begin{question}
 	What is the impact of  platform's  incomplete information about workers on the platform's incentive mechanism and cost, compared with complete information?
\end{question}

Third, the platform needs to   optimize the load assignment and recovery threshold, which  will affect the determination of   incentives as well as the performance of coded machine learning. 
If  the recovery threshold is too small,   the  load of each worker becomes so large that the overall runtime   will eventually increase. If the recovery threshold is too large, the   redundancy may not be sufficient to mitigate the effect of straggling workers. 
This   leads to the third key question of this paper: 
\begin{question}
	How should the platform design the load assignment and recovery threshold     for the incentivized workers?
\end{question}



%

\subsection{Contributions}

We summarize our key novelty and   contributions below.
\begin{itemize}
	\item \emph{Incentive mechanisms   for coded machine learning:} To the best of our knowledge, this is  the first analytical work to study economic   incentives  in coded machine learning. 
	The incentive issue determines whether coded machine learning can be implemented with enough workers'  participation. 
	
	\item \emph{Easy-to-implement incentive mechanisms for heterogeneous workers:} 	We perform the analysis considering  workers' multi-dimensional heterogeneity in computation  costs and performances under incomplete information  and stochastic  information.	As a result, searching  the target incentivized workers  among $M$-many types of workers would  require  large-scale computation. We	manage to summarize workers’ multi-dimensional heterogeneity into a one-dimensional 	metric to guide the platform's worker selection under incomplete information. 
	Such a metric  reduces the computation complexity   from $\mathcal{O}(2^M)$   to  $\mathcal{O}(M)$.
	We only need  to set the  recovery threshold   linearly  proportional to the total participator number when using MDS codes, which is easy to implement. 	
	
	\item \emph{Impact of incomplete information for mechanism design:} 
	We show that the platform may give smaller rewards to the workers with efficient computation performances, compared with workers with poor performances under complete information. However, the platform  gives larger rewards to the workers with efficient computation performances  under incomplete information. These reward-performance relationships are independent of the workers' distribution in each type. We also show 
	that the platform's increased cost due to incomplete information   disappears when    worker number is sufficiently large, but it does not monotonically decrease in worker number.
\end{itemize}

\vspace{-1.5mm}
\subsection{Related Work}
\vspace{-0.5mm}
\label{literature}

Most literature about coded machine learning  focused on   designing  coding schemes to mitigate stragglers and improve efficiency in distributed computation, e.g., MDS codes \cite{lee2017speeding}, short-dot codes \cite{dutta2016short}, polynomial codes \cite{yu2017polynomial}, and s-diagonal codes \cite{wang2019computation}. These studies usually focused on the  matrix computation tasks such as matrix-vector multiplication (e.g., \cite{reisizadeh2017latency}),  matrix-matrix multiplication (e.g., \cite{lee2017high}), and large-sparsity  matrix computation (e.g., \cite{suh2017matrix}). 
Moreover, a few studies further considered the secure coded computation problem (e.g., \cite{bitar2017minimizing}) and the computation performance heterogeneity of   workers (e.g., \cite{reisizadeh2019coded}). 
 
However, existing studies make  an optimistic assumption that workers are willing to participate in  coded machine learning, which may not be realistic without proper incentives to the workers. To the best of  our knowledge, this paper is the first attempt that analytically studies the platform's incentive mechanism design in coded machine learning.


The rest of the paper is organized as follows. The system model is described in Section \ref{model}. We study the platform's   incentive mechanisms under complete information and  incomplete information in Sections \ref{complete} and \ref{incomple}, respectively.   We perform  simulations in Section \ref{simulation} and   conclude in Section  \ref{conclusion}.

\section{System Model and Preliminaries }
\label{model}
We consider a distributed  coded machine learning system  involving a platform and $N$ workers. 
In the following, we will first introduce the   platform's task coding and load assignment  as well as  workers' heterogeneity, and then specify the information scenarios, and finally  formulate the   Stackelberg game between workers and the platform. 

\subsection{Platform Modeling: Task Coding and Load Assignment }
\label{2a}
In this subsection, we    introduce how   the platform encodes the original task and assigns coded subtasks to workers.

This paper focuses on  matrix-vector multiplication tasks in the widely adopted problems such as  gradient descent-based algorithms, logistic regression, and reinforcement learning  (e.g. \cite{reisizadeh2019coded,reisizadeh2017latency})\footnote{Note that our analysis and insights can be easily applied to matrix-matrix multiplications (which can be decomposed to several  matrix-vector multiplications) and other machine learning tasks as long as we can specify the relationship between workers' computation loads and computation time.}.  Given a data matrix $\mathbf{A} \in \mathbb{R}^{r \times s}$, the platform wants to compute the task $\mathbf{y}=\mathbf{A} \mathbf{x}$ for an input vector $\mathbf{x} \in \mathbb{R}^{s}$.

Regarding the task coding, we use  linear combinations of the $r$ rows of the matrix $\mathbf{A}$ to generate the computation redundancy, such that the platform  can recover the result $\mathbf{Ax}$ as soon as  receiving any $r$ inner products from the workers.
We take $(n, k)$-MDS codes as an example. If the task is distributed across $n$ workers, for any $ k \in \{1,...,n\}$, the platform first divides  matrix $\mathbf{A}$  into $k$ equal-sized submatrices in $\mathbb{R}^{\frac{r}{k} \times s}$. Then, by applying an $(n, k)$-MDS code, the platform obtains $n$ encoded submatrices with unchanged size ${\frac{r}{k} \times s}$, one  for each worker.  Upon receiving any $k$ workers' results, the platform can  decode the result of the  original task. 

The  \emph {computation load} of a worker's subtask is defined as the number of inner products of  assigned coded rows of $\mathbf{A}$ with $\mathbf{x}$.  
If worker $i$ is assigned a matrix-vector multiplication with matrix size $\ell_{i}\times s$,  
his computation load is  $\ell_{i}$. 
In the example of   $(n, k)$-MDS codes,   each worker has the same computation load $\ell={r}/{k}$.



Given the computation task, the platform needs to assign appropriate   subtasks   for heterogeneous workers.



\subsection{Worker Modeling: Heterogeneous Costs and   Performances}
We consider a population of $N$ workers in the coded machine learning system, with heterogeneous computation performances and costs as modeled below.

\subsubsection{Computation Performances}
We  consider  a 2-parameter shifted exponential distribution for the computation time of each worker, which is widely used in literature  (e.g., \cite{reisizadeh2019coded,reisizadeh2017latency}) because of the  good approximation to the experiment statistics. After being assigned $\ell_i$ rows of the matrix-vector multiplication, worker $i\in \{1,...N\}$ will finish the computation subtask in time $T_i$, which is a random variable with the following cumulative distribution function (CDF):
\begin{equation}
\label{dist}
\begin{split}
\operatorname{Pr}\left(T_{i} \leq t\right)=&1-e^{-{\mu_i}\left(\frac{t}{\ell_i}-a_i\right)}, \forall t \ge a_i\ell_i,
\end{split}
\end{equation}
where parameter $\mu_i > 0$  is worker $i$'s average computation speed and parameter  $a_i>0$   tells worker $i$'s start-up time to begin the computation. 
Equation \eqref{dist} shows that a worker with a larger $\mu_i$, a smaller $a_i$, or a smaller $\ell_{i}$ is more likely to finish his computation earlier. 

We   consider that each worker's computation time distribution  in \eqref{dist} (including the form as well as parameters $\mu_i$ and $a_i$) is known by the platform and   other workers in this paper. However, its realized   value  is unknown to both the platform and the worker himself ahead of time,  due to   computing resources' unpredictable  noise nature\footnote{The joint consideration of heterogeneous stochastic information and the  private information (i.e., computation costs to be introduced below) is a key feature of our model. It differs from many incentive mechanism design papers for networking problems, where the workers have no information uncertainty regarding their own costs or performances (e.g., \cite{ningwiopt,ma2018incentivizing}).}.  A worker will return the computation results to the platform  as soon as he finishes  his computation. Thus, each worker's realized computation performance is known to the platform after computation.

\subsubsection{Computation Costs}
The computation may lead to many costs, such as  time cost, energy consumption, and  the potential negative impact to other applications. We  focus on the time cost in this paper, as the  commodity servers usually charge workers on time (e.g. Amazon EC2 \cite{ec2}). 
We consider each worker $i$'s  computation cost  as $c_i$ per unit time of computation.

\subsubsection{Worker Types}
Given workers' multi-dimensional heterogeneity in computation performances and costs,  we introduce \emph{worker type} to classify the large number of  workers. 
Workers   are distinguished by  the   cost, average computation speed, and the start-up computation time. 
We  call a worker with $(c_m, \mu_m,a_m)$ as a type-$m$ worker. All $N$ workers belong to a set $\mathcal{M}=\{1,2,...,M\}$ of $M$ types. Each type $m\in \mathcal{M}$ has $N_m$ workers, with $\sum_{m=1}^{M}N_m=N$. We will see later in Section \ref{s1} that workers' three-dimensional heterogeneity greatly increases the complexity and difficulty of platform's incentive mechanism design. 


\vspace{-2mm}
\subsection{Information Scenarios}
\vspace{-1mm}
To study the impact of incomplete information, we will perform the analysis in different information scenarios.

As for workers' computation time, the platform   only knows  each worker's   computation time distribution (i.e., $\mu$ and $a$). Regarding workers' computation costs,   we will consider  two information cases  in the following:
\vspace{-1mm}
\begin{enumerate}
	\item   \emph{Complete information (benchmark)}: The platform knows each worker's  computation cost $c$  (and thus his type).  This may not be easy for  the platform to achieve,  but  it provides   the platform's minimum cost in all information cases for comparison. 
	\item \emph{Incomplete information}: The platform knows the total number of workers $N$ and the specific number of each worker type $\{N_m\}_{m\in\mathcal{M}}$, but does not know each worker's  computation cost $c$ (and thus his type).
\end{enumerate}
\vspace{-1mm}

Workers always have incomplete information about each other's costs but know the computation time distributions.

\vspace{-2mm}
\subsection{Two-Stage Stackelberg Game Formulation}
\vspace{-1mm}
In this subsection, we specify the strategies, payoffs, as well as costs of the platform and workers.

\subsubsection{Strategies of Platform and Workers }
\begin{figure}[tbp]
 	\centering
 	\includegraphics[width=1\linewidth]{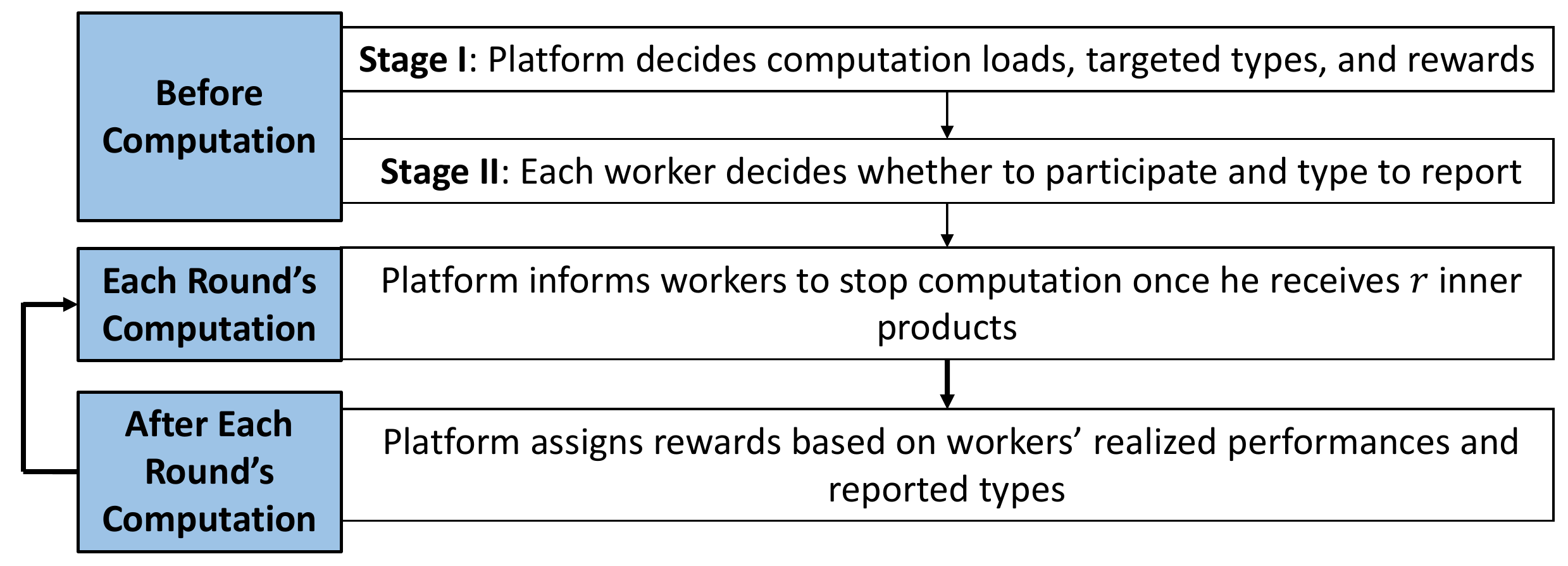}
 	\vspace{-5mm}
 	\caption{Workflow of coded machine learning with incentives:   the  platform and workers reach an agreement by playing a Stackelberg game before computation,  workers perform each round's computation, and the platform rewards workers after each round's computation.}
 	\label{fig16}
 	\vspace{-5mm}
 \end{figure}

As shown in Fig.~\ref{fig16}, we model the decision making of workers and the platform before the computation as a two-stage Stackelberg game. 
\begin{itemize}
	\item In Stage I, the  platform announces  the set of targeted worker types   $\mathcal{S}\subseteq \mathcal{M}$,  the computation loads in each computation round   $\boldsymbol{\ell}\triangleq \{\ell_{m}\}_{m\in \mathcal{S}}$, and the rewards in each round $\{p_m^j\}_{m\in \mathcal{M},j\in \{1,...,k\}}$, where $p_m^j$ is the reward for a type-$m$ worker being the $j$-th to finish  the computation.  Thus, $p_m\triangleq\sum_{j=1}^{k}Pr_m^jp_m^j$ is the expected reward for a type-$m$ worker in each round, where $Pr_m^j$ is the probability of a type-$m$ worker being the $j$-th  to finish the computation\footnote{The platform and workers know   the participating worker types (i.e.,  targeted types in $\mathcal{S}$), the worker number of each type $N_m$,  and workers' computation time distributions in \eqref{dist} under both complete and incomplete information, so both the platform and workers can derive $Pr_m^j$.}. Equivalently, the platform actually determines the expected rewards for workers $\boldsymbol{p }\triangleq \{p_m\}_{m\in \mathcal{M}}$.

\item In Stage II, after knowing  the platform's decisions,  each    worker decides  whether to participate and what type to report (which may not be his true type), as the platform will give rewards based on workers' realized performances and their reported types after the computation.
\end{itemize}
The coded machine learning applies to multiple computation rounds (e.g. in gradient descent problem). 
 In each round of  computation, once the platform receives a decodable subset
of computation results (i.e., $r$  inner products), the platform will inform other workers to stop the computation without any waste \cite{aktas2017effective}. After each round's computation, the platform rewards    workers as announced in Stage I.

 Note that the number of participating workers $n=N^{(\mathcal{S})}\triangleq\sum_{m \in \mathcal{S}}N_m$ is controlled by the incentive mechanism, and the recovery threshold $k$    depends on the load assignment as illustrated in Section \ref{2a}.

\subsubsection{Workers' Payoff}
Due to the randomness of the computation time, each worker can only maximize his expected payoff. 
Each worker's expected payoff in each round is the difference between  his expected  reward and  expected computation cost.  

If a type-$m$ worker  optimally reports/misreports himself as type $\tilde{m}$, his expected reward in each round is $p_{\tilde{m}}$. 

In each round, all workers' expected spent time  is the expected overall runtime denoted by $\mathbb{E}[T]$. 
Note that even if a worker finishes earlier in a particular round, he needs to wait some random time for the rest of workers to finish and cannot turn to any other job, resulting in the same time occupation   for all the workers.  Thus, a type-$m$ worker's expected computation cost is $c_m\mathbb{E}[T ]$.


In summary, if a type-$m$ worker participates and  reports  himself as type $\tilde{m}$, his expected payoff in each round is:
\begin{equation}
\label{uhet}
\begin{split}
&\mathbb{E}[U (c_m,\mu_m,a_m,{\tilde{m} })]=p_{\tilde{m}}-c_m\mathbb{E}[T ].\\
\end{split}
\end{equation}
Workers can manipulate his type reporting to achieve the maximum payoff. Each worker will participate in the computation once he expects a non-negative payoff in \eqref{uhet}.


\subsubsection{Platform's Cost}
The platform's expected cost objective in each round  involves the overall runtime  and total payment to   targeted workers. If the platform considers workers' incentive compatibility in designing the incentives (as   guaranteed in Section \ref{incomple}),  all   workers will truthfully report their types   and   the platform's expected cost in each round is 
\begin{equation}
\label{whet}
\begin{split}
&\mathbb{E}[W (\boldsymbol{p },\mathcal{S},\boldsymbol{\ell })]
=\gamma_1\mathbb{E}[T ]+\gamma_2\sum_{m\in \mathcal{S}}N_mp_m,\\
\end{split}
\end{equation}
where term $\mathbb{E}[T ]$ is the expected overall runtime   and term $\sum_{m\in \mathcal{S}}N_mp_m$ is the expected  total payment to targeted worker types in set $\mathcal{S}$. 
Terms $\gamma_1$ and $\gamma_2$ indicate the platform's valuations on expected overall runtime and expected payment, respectively. 


\begin{table}[tbp]
	\caption{Key Notations}  
	\vspace{-4.5mm}
	\begin{center}
		\begin{tabular}{|c|c|}
			\hline
			$N$ &  Number of workers \\
			\hline
			$M$ &  Number of worker types \\

			\hline
			$r$ &  Number of rows of matrix $\mathbf{A}$ or simply total workload \\
			\hline
			$\mu_m$ &    Average computation speed  of type-$m$ workers \\
			\hline
			$a_m$ &   Start-up time  of type-$m$ workers\\
			\hline
			$c_m$ & Unit cost of type-$m$ workers\\
			\hline
			$p_m$& Expected reward for a  type-$m$ worker\\
			\hline
			$\ell_m$ & Computation load in term of rows  for a  type-$m$ worker\\
			\hline
			$\mathcal{S}$ & Set of worker types that platform targets at\\
							\hline
			$N^{(\mathcal{S})}$ &  Number of participating workers \\
			\hline
			$T_i$& Worker $i$'s random computation time\\
			\hline
			$\mathbb{E}[T]$ & Expected overall runtime\\
			\hline
		\end{tabular}
		\label{tablit} 
		\vspace{-5mm}
	\end{center}
\end{table}
\begin{table}[tbp]
	\caption{Information Scenarios}  
	\vspace{-4.5mm}
	\begin{center}
		\begin{tabular}{|c|c|c|}
			\hline
			\textbf{  Scenarios}&\textbf{Worker  heterogeneity} &  \textbf{Platform's knowledge} \\
			\hline
			Complete-Hetero & \tabincell{c}{Computation costs \\ and performances } &  \tabincell{c}{Costs and  computation\\ time distributions} \\
			\hline
			Incomplete-Hetero & \tabincell{c}{Computation costs\\ and performances} & \tabincell{c}{ Computation \\time distributions}\\
						\hline
			\tabincell{c}{Incomplete-\\HeteCostOnly} & Computation costs&  \tabincell{c}{ Computation \\time distribution} \\
			\hline
		\end{tabular}
		\label{sce} 
		\vspace{-6mm}
	\end{center}
\end{table}

For ease of reading, we list the key notations in Table \ref{tablit} and all information scenarios that we will study in Table \ref{sce}. In the following, we will   study the platform's   incentive mechanism design under complete information  (i.e., ``Complete-Hetero'' scenario)   in Section \ref{complete}  and that    under incomplete information (i.e., ``Incomplete-Hetero'' scenario) in Section \ref{hetero}.  Moreover, we   will consider  a special case with workers' heterogeneity only in costs  (i.e., ``Incomplete-HeteCostOnly''  scenario) in Section \ref{homo} to further explore the optimal load assignment and recovery threshold in coded machine learning with incentives.

\section{Incentive Mechanism Design under Complete Information}
\label{complete}
In this section, we study the platform's optimal incentive mechanism  under the assumption that the platform   knows each worker's type (i.e., ``Complete-Hetero'' scenario). Although    complete information   may not be practical for the platform to achieve,  it serves as a benchmark for later comparison with incomplete information. 

\subsection{Computation Load Assignment and Overall Runtime}
We  first present the load assignment scheme  and compute the corresponding overall runtime.

With workers' heterogeneous computation performances,  assigning equal loads to all workers   (e.g., in the MDS coding scheme) is clearly not optimal.  It is challenging to compute the optimal load assignment in this case, because   the expected overall runtime objective is difficult to derive under nonuniform load assignment  
due to   uncertain order of finish workers and workers' heterogeneous computation time distributions.

To simplify the analysis, we  apply  asymptotic analysis as the number of workers   goes to infinity, similar to \cite{reisizadeh2019coded} which  focuses on minimizing the expected overall runtime. The asymptotically optimal load assignment and the corresponding expected overall runtime are given in Lemma \ref{lm1}:
\begin{lemma} 
	\label{lm1}
	When worker number  goes to infinity,  the following computation loads minimize  the expected overall runtime: 
\begin{equation}
\label{4}
\begin{aligned}
\ell_{m} &=\frac{r}{ \lambda_{m}\sum_{m\in \mathcal{S}} N_m\frac{\mu_{m}}{1+\mu_{m} \lambda_{m}}},\forall m\in \mathcal{S},
\end{aligned}
\end{equation}
where  $\lambda_{m}$ is the unique positive solution to  $e^{\mu_{m} (\lambda_{m}-a_{m})}=\mu_{m} \lambda_{m}+1$. 
The   expected overall runtime under  scheme \eqref{4} is 
\begin{equation}
\label{time1}
\mathbb{E}[T]=\frac{r}{\sum_{m\in \mathcal{S}} N_m\frac{\mu_{m}}{1+\mu_{m} \lambda_{m}}}.
\end{equation}
\end{lemma}
Proof of Lemma \ref{lm1} is given in Appendix I of the technical report \cite{appen1}. 
Term $\lambda_{m}$ is introduced for representing the closed form  of the minimum point. The $1/\lambda_{m}$ indicates the computation performance of type-$m$ workers, as  $\frac{\partial\lambda_{m}}{\partial\mu_{m}}\le 0$ and $\frac{\partial\lambda_{m}}{\partial a_{m}}\ge 0$. We assign a larger computation load to a worker   with better computation performance. 
The assignment scheme \eqref{4} can be viewed as an approximation of  the finite $N$ case with approximation error $o(1)$ and is applicable to both ''Complete-Hetero'' and ''Incomplete-Hetero''   scenarios.

Next, we use backward induction to  study   workers'    decisions  in Stage II  and   the platform's   strategies in Stage I.

\subsection{Workers'   Participation Decisions in Stage II}
Based on  the  load assignment in \eqref{4}, the  rewards, and the targeted  worker type set $\mathcal{S}$ announced to   workers by the platform, each   worker   decides whether to participate in the computation. Note that here the platform knows each worker's type under complete information, so workers cannot misreport their types.

According to \eqref{uhet} and \eqref{time1}, a type-$m$ worker's expected payoff when truthfully reporting (i.e., $\tilde{m}=m$) and handling the assigned load   $\ell_m$ in \eqref{4} is:
\begin{equation}
\label{7}
\begin{split}
\mathbb{E}[U (c_m,\mu_m,a_m,m)]=p_m-{c}_m\frac{r}{\sum_{m\in \mathcal{S}} N_m\frac{\mu_{m}}{1+\mu_{m} \lambda_{m}}}.
\end{split}
\end{equation}

Each worker locally decides to participate as long as he obtains a non-negative expected payoff  in \eqref{7}.

\subsection{Platform's Worker Selection and Rewards   in Stage I}
\label{s1}
Considering workers'    decisions, the platform determines  the targeted  worker type set and  the  rewards  for   workers  in Stage I,   by balancing workers' computation performances and costs.  

\subsubsection{Problem Formulation}
According to \eqref{whet} and \eqref{time1}, the platform's expected cost objective is 
\begin{equation}
\label{8}
\begin{split}
\mathbb{E}[W (\boldsymbol{p },\mathcal{S})]=\gamma_1\frac{r}{\sum_{m\in \mathcal{S}} N_m\frac{\mu_{m}}{1+\mu_{m} \lambda_{m}}}+\gamma_2\sum_{m \in \mathcal{S}}N_mp_m.\\
\end{split}
\end{equation}

Under complete information,   the platform only needs to ensure that  targeted workers obtain non-negative payoff \eqref{7}, i.e.,  satisfy   Individual Rationality (IR) constraints as below:  
\begin{defn}[Individual Rationality]
	The incentive mechanism is individually rational if each targeted type-$m\in \mathcal{S}$ worker receives a non-negative payoff 	by accepting the expected reward $ {p_m }$ intended for his type, i.e.,
	\begin{equation}
	\label{9}
	\mathbb{E}[U (c_m,\mu_m,a_m,m)]\ge 0, \forall m \in \mathcal{S}.
	\end{equation}
\end{defn}

The optimization problem of the platform in the  ``Complete-Hetero'' scenario is formally given below.
\begin{problem}[Complete-Hetero]
	\label{c}
	\begin{equation*}
	\begin{split}
	\min \quad&\mathbb{E}[W (\boldsymbol{p },\mathcal{S})]\\
	s.t.\quad &\mathbb{E}[U (c_m,\mu_m,a_m,m)]\ge 0, \forall m \in \mathcal{S}\\
	var. \quad & {\boldsymbol{p }\in [0,\infty)^M,\mathcal{S}\subseteq\mathcal{M}}\\ 
	\end{split}
	\end{equation*}
\end{problem}

To minimize the cost objective, the platform   sets each targeted worker's expected reward just enough to cover his expected cost, i.e., $\mathbb{E}[U (c_m,\mu_m,a_m,m)]= 0, \forall m \in \mathcal{S}$. By substituting the derived $p_m$ into \eqref{8}, we can simplify Problem \ref{c} to only choose the optimal  worker types to target as follows.
\begin{problem}[Worker Selection under Complete Information]
	\label{aa}
	\begin{equation*}
\begin{split}
\min_{\mathcal{S}\subseteq \mathcal{M}}\left(\gamma_1+\gamma_2\sum_{m\in \mathcal{S}}N_m{c}_m\right)\frac{r}{\sum_{m\in \mathcal{S}} N_m\frac{\mu_{m}}{1+\mu_{m} \lambda_{m}}}.\\
\end{split}
\end{equation*}
\end{problem}
Problem \ref{aa} is a combinatorial
optimization problem and should be solved by balancing workers' computation performances and costs. It is challenging to directly find  the optimal subset of workers given three-dimensional heterogeneity (i.e., $\mu_m$, $a_m$, and $c_m$), which in general requires combinatorial search over all $\sum_{n=1}^{M}C_M^n=2^M-1$ possible subsets of worker types.  The  complexity is $\mathcal{O}(2^M)$,  which is infeasible for a large number   of worker types. Alternatively, we will   explore to summarize the  multi-dimensional heterogeneity as a single metric to guide the platform's worker selection in linear complexity.

\subsubsection{Optimal Solution}
We will use two steps to summarize the three-dimensional heterogeneity (i.e., $\mu_m$, $a_m$, and $c_m$) into a one-dimensional metric. 
We first summarize the two-dimensional heterogeneity of workers' performances (i.e., $\mu_m$ and $a_m$) as  
\begin{equation}
\phi_m\triangleq\frac{\mu_{m}}{1+\mu_{m} \lambda_{m}}.
\end{equation}
The $\phi_m$ can represent type-$m$ workers' computation performance as it increases in average speed $\mu_m$ and decreases in start-up time $a_m$ (as $\lambda_{m}$ decreases in $\mu_m$ and increases in $a_m$). 
We further summarize the   cost $c_m$ and the performance $\phi_m$  as
\begin{equation}
\Omega_m \triangleq \frac{c_m}{\phi_m}.
\end{equation}
The $\Omega_m$ is type-$m$ workers' \emph{cost-performance ratio}. Without loss of generality, we assume increasing  cost-performance ratios   among all the $M$ worker types:  $\Omega_1\le ...\le   \Omega_M$.

The following Theorem \ref{thmc}  shows how the one-dimensional metric $\Omega$ guides the platform's selection in the presence of workers' three-dimensional   information. 
\begin{theorem}
	\label{thmc}
	In the ``Complete-Hetero'' scenario,  there exists a threshold type $n^{ComHet}$,
	\begin{equation}
	\label{12}
	n^{ComHet}\hspace{-1mm} = \hspace{-1mm} \max\left\{n|n\in \mathcal{M},\frac{c_n}{\phi_n}\le \frac{\gamma_1+\gamma_2\sum_{m=1}^{n}N_mc_m}{\gamma_2\sum_{m=1}^{n}N_m\frac{\mu_{m}}{1+\mu_{m} \lambda_{m}}}\right\},
	\end{equation}
	such that it is optimal for the platform to induce the participation of worker types in the following targeted set $\mathcal{S}^{ComHet}$:
		\begin{equation}
	\begin{split}
	\mathcal{S}^{ComHet}  =\{n|n\in \mathcal{M},n\leq n^{ComHet}\}.\\
	\end{split}
	\end{equation} 
 The   optimal expected reward  for a type-$m$ worker is
	\begin{equation}
	\begin{cases}
	&\hspace{-3mm}p_m^{ComHet}={c}_m\frac{r}{\sum_{m\in \mathcal{S}^{ComHet}  } N_m\frac{\mu_{m}}{1+\mu_{m} \lambda_{m}}}, \; \forall m \in \mathcal{S}^{ComHet}  ,\\
	&\hspace{-3mm}p_m^{ComHet}=0, \;\forall   m \notin \mathcal{S}^{ComHet}  .
	\end{cases}
	\end{equation}
\end{theorem}
\begin{figure}[tbp]
	\centering
	\includegraphics[width=0.5\linewidth]{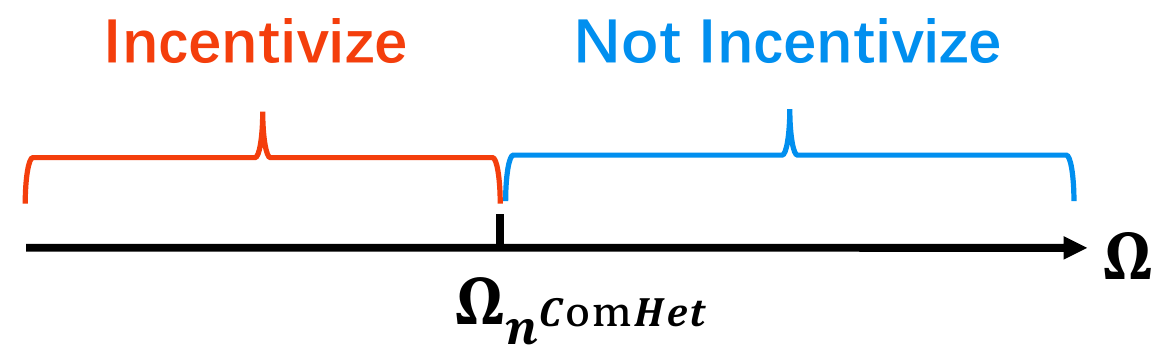}
	\vspace{-2.5mm}
	\caption{Illustration of $\mathcal{S}^{ComHet}$ in ``Complete-Hetero'' scenario.}
	\label{figb}
	\vspace{-5mm}
\end{figure}
Proof of Theorem \ref{thmc} is given in Appendix II of the technical report \cite{appen1}. 

As illustrated in Fig.~\ref{figb}, Theorem \ref{thmc}  shows that with workers multi-dimensional  heterogeneity in computation performances and costs, the platform  will incentivize the worker types with a small cost-performance ratio under complete information. In other words, the platform has a trade-off between workers' computation performances and  computation costs. The platform can tolerate a larger cost $c_m$ when the computation performance is better (i.e., faster average computation speed $\mu_m$  or less start-up time $a_m$).  

For the worker types that the platform chooses to incentivize (i.e., types in $\mathcal{S}^{ComHet}  $), the platform precisely makes the expected rewards  just enough to  compensate for each worker's expected computation cost under complete information. 
For the worker types  out of targeted set $\mathcal{S}^{ComHet}  $, the platform will not incentivize them by setting a zero expected reward. We will see in next section that it is impossible for the platform to  make all  participating workers    obtain a zero expected payoff under incomplete information, unless he only targets one type.

Moreover, we have the following remark about the complexity of this easy-to-implement approach.
\begin{remark}
	The platform can compute $\mathcal{S}^{ComHet}$   with a linear complexity $\mathcal{O}(M)$, which is much more efficient than $\mathcal{O}(2^M)$.
\end{remark}

%

The complete information scenario  servers as  a benchmark to help understand the more realistic case of incomplete information scenario, which we will study in next section. 

\section{Incentive Mechanism Design under Incomplete Information}
\label{incomple}
In this section, we study the optimal incentive mechanism in the more realistic incomplete information scenario. 
Moreover,   we will  look at a special case   where workers have homogeneous computation time distribution  (e.g., under the same equipment or configuration)  but still   different   costs in Section \ref{homo}, to  further study the optimal load assignment and recovery threshold under incentives.

\subsection{Workers' Heterogeneity in Both  Performances and Costs}
\label{hetero}
In this subsection, we focus on the general ''Incomplete-Hetero'' scenario where workers' heterogeneity lies in both computation performances and computation costs, i.e., workers have different $\mu_{i}$, $a_i$, and $c_i$. 
The platform does not know each worker's marginal cost, but he knows what types there are, the number of each worker type, and   each  worker's computation time distribution. 

\subsubsection{Workers'   Participation Decisions in Stage II}
Based on  the  load assignment in \eqref{4},   the rewards, and the   targeted  worker type set $\mathcal{S}$ announced to   workers by the platform, each   worker decides whether to participate and (if yes) which type to report\footnote{When a worker decides to participate, he can only choose to report a type that is within the targeted worker type set announced by the platform.}.

According to \eqref{uhet} and \eqref{time1}, if  a type-$m$ worker  participates and reports his type as $\tilde{m}$, his expected payoff in this case is\footnote{Note that misreport   will not affect   the expected overall runtime. After  task  coding, the number of subtasks for workers is fixed, and the platform knows workers' $\mu$ and $a$ which cannot be misreported. Thus, under incomplete information, even if some workers with types not in $\mathcal{S}$ want  to participate through misreporting their cost types,   the number of   workers who can perform the computation and these workers' computation performances  will not change.   }:
\begin{equation}
\label{18}
\begin{split}
\mathbb{E}[U (c_m,\mu_m,a_m,\tilde{m})]\hspace{-1mm}=p_{\tilde{m}}-{c}_m\frac{r}{\sum_{m\in \mathcal{S}} N_m\frac{\mu_{m}}{1+\mu_{m} \lambda_{m}}}.
\end{split}
\end{equation}
If a worker does not participate, his expected payoff is 0.

Each   worker will participate as long as he can obtain a non-negative payoff, and he will report the type which maximizes his expected payoff to the platform.  
Thus, a type-$m \in \mathcal{M}$ worker's optimal reported type (if he participates)  is 
\begin{equation}
\label{17}
\tilde{m}^*=\arg\max_{\tilde{m}}\mathbb{E}[U (c_m,\mu_m,a_m,\tilde{m})], 
\end{equation}
and a type-$m \in \mathcal{M}$ worker's optimal participation decision is 
\begin{equation}
\label{18a}
\begin{cases}
&\mathrm{Participate, \hspace{7.5mm}if }\;\mathbb{E}[U (c_m,\mu_m,a_m,\tilde{m}^*)]\ge 0,\\
&\mathrm{Not \; participate,\; if }\;\mathbb{E}[U (c_m,\mu_m,a_m,\tilde{m}^*)]< 0.\\
\end{cases}
\end{equation}


\subsubsection{Platform's Strategies in Stage I}
Taking  workers' decisions into consideration, the platform   determines the optimal targeted  worker type set and  the optimal   rewards for workers.

Since the platform does not know each worker's type, the platform   needs to ensure a non-negative payoff for each targeted worker type and make sure that all   workers do not misreport their cost types\footnote{Revelation principle demonstrates that if a social choice function can be implemented by an arbitrary mechanism, then the same function can be implemented by an incentive-compatible-direct-mechanism (i.e. in which workers truthfully report   types)  with the same equilibrium outcome. Thus, requiring IC will simplify the mechanism design without affecting the optimality.}. In other words,  except for the IR constraints, the platform has to make decisions under  the Incentive Compatibility (IC) constraints:

\begin{defn}[Incentive Compatibility]
	Then incentive mechanism is incentive 	 compatible if each type-$m\in \mathcal{M}$ worker maximizes his own payoff by truthfully reporting his type, i.e.,
	\begin{equation}
	\label{IC}
	\mathbb{E}[U (c_m,\mu_m,a_m,m)]  \ge  \mathbb{E}[U (c_m,\mu_m,a_m,{\tilde{m} })],   \forall m, \tilde{m}  \in \hspace{-0.6mm} \mathcal{M}.
	\end{equation}
\end{defn}

Formally, the platform's optimization problem in ``Incomplete-Hetero'' scenario is:
\begin{problem}[Incomplete-Hetero]
	\label{d}
	\begin{equation*}
	\begin{split}
	\min \quad&\mathbb{E}[W (\boldsymbol{p },\mathcal{S})]\\
	s.t.	\quad & \rm{IR}\; \rm{Constaints\; in \; } \eqref{9},\; \rm{IC}\; \rm{Constaints\; in \; }\eqref{IC}\\
	var. \quad& {\boldsymbol{p }\in [0,\infty)^M,\mathcal{S}\subseteq\mathcal{M}} \\
	\end{split}
	\end{equation*}
\end{problem}


Compared with Problem \ref{c}, the incomplete information further increases the complexity and difficulty of solving Problem \ref{d} through the additional $M(M-1)$ IC constraints. We present the platform's optimal strategies and the corresponding proof sketch as follows: 
\begin{theorem}
	\label{thmd}
	In the ``Incomplete-Hetero'' scenario,  there exists a threshold type $n^{IncHet}$,
	\begin{equation}
	\label{23}
	n^{IncHet}=\arg\min_{n\in \mathcal{M}}\frac{\gamma_1+\gamma_2\frac{c_n}{\phi_n}\sum_{m=1}^nN_m\phi_m}{\sum_{m=1}^n N_m\phi_m},
	\end{equation}
	such that it is optimal for the platform to induce the participation of worker types in the following targeted set $\mathcal{S}^{IncHet}  $:
	\begin{equation}
	\begin{split}
	\mathcal{S}^{IncHet}  =\{n|n\in \mathcal{M},n\leq n^{IncHet}\}.
	\end{split}
	\end{equation} 
	The platform's optimal expected reward  for a  type-$m\in \mathcal{M}$ worker is
	\begin{equation}
	\label{24}
	p_m^{IncHet}=\phi_m\frac{rc_{n^{IncHet}}}{\phi_{n^{IncHet}}\sum_{m\in \mathcal{S}^{IncHet}  } N_m\frac{\mu_{m}}{1+\mu_{m} \lambda_{m}}}, \forall m\in \mathcal{M}.
	\end{equation}
\end{theorem}
Proof of Theorem  \ref{thmd} is given in Appendix III of the technical report \cite{appen1}. 

Theorem \ref{thmd}  shows that the platform is able to derive the optimal set of workers under incomplete information in linear complexity, by  summarizing the three-dimensional type information  into a   one-dimensional metric (i.e.,  cost-performance ratio $\Omega$). 
Although the platform uses the same metric $\Omega$ to select workers in scenarios ``Incomplete-Hetero''  and ``Complete-Hetero'', the numbers of targeted types (i.e., $n^{ComHet}$ in \eqref{12} and $n^{IncHet}$ in \eqref{23}) are different. 
One may wonder whether the platform   always targets fewer worker types  under incomplete information. Our simulation results in Section \ref{simulation} show that it is not always true. 
In the two information scenarios, different payment rules will have different influences on the   platform's expected cost, resulting in  different worker selection rules in \eqref{12} and \eqref{23}.

Different from the ``Complete-Hetero'' scenario where all participating workers obtain a zero payoff,   in the ``Incomplete-Hetero'' scenario, all worker types in set $\mathcal{S}^{IncHet}$ except the boundary type obtain positive expected payoffs which are the \emph{information rent} in economics.  In the ``Complete-Hetero'' scenario, the platform may give smaller expected rewards to the workers with efficient computation performances, compared with workers with poor performances, as the platform knows workers' costs. However,   in ``Incomplete-Hetero'' scenario, the platform   sets a larger  expected reward  for a type with better computation performance $\phi$ as shown in \eqref{24}. Such a reward-performance relationship is independent of the workers' distribution in each type. Moreover, the rewards in Theorem \ref{thmd} ensure that worker types in the desirable set $\mathcal{S}^{IncHet}$ will participate, and 
types not  in $\mathcal{S}^{IncHet}$ will  not participate as the expected reward cannot compensate for their expected costs.  

Furthermore, we have the following  insight about the optimal number of targeted   types in the two scenarios:
\begin{corollary}
	\label{pro}
	The number of targeted types increases in  the platform's valuation on expected overall runtime $\gamma_1$ and decreases in the platform's valuation on payment $\gamma_2$.
\end{corollary}
Proof of Corollary \ref{pro} is given in Appendix IV of the technical report \cite{appen1}.  
If the platform attaches more attention on the computation time (larger $\gamma_1$), he is
likely to include more worker types to reduce the overall runtime; if the platform pays more attention on the payment to workers (larger $\gamma_2$), he will incentivize fewer worker types to save   money.

Note that the recovery threshold $k$  is a random variable in these two scenarios.  Workers' different loads and random computation time   lead to the randomness of the number of  workers who finish the computation before the platform receives $r$ rows (i.e., the value of $k$ in $\sum_{j=1}^{k}\ell^j=r$, where $\ell^j$ is the load of the $j$-th worker who finishes the computation). 
Next, we will consider a special case where workers have the same computation time distribution but different costs to further study the optimal load assignment and recovery threshold in coded machine learning with incentives.

\subsection{Workers' Heterogeneity  in   Costs Only}
\label{homo}
In this subsection, we focus on the ``Incomplete-HeteCostOnly'' scenario where workers have heterogeneous  computation costs $c$ and  the same computation performances (i.e.,   computation time distribution).  The platform has incomplete information about workers' costs but knows the  workers' computation time distribution and worker type distribution. 
 
Each worker $i$'s computation time (i.e.,  $T_i$) in this case follows \eqref{dist} with   $\mu_{i}=\mu$, $a_i=a$, and $\ell_{i}=\ell, \forall i \in \{1,...,N\}$.
In this case, we consider the   $(n, k)$-MDS code as introduced before, which is  a widely used code for workers with homogeneous computation time distribution in  literature (e.g.,  \cite{lee2017speeding,reisizadeh2017latency}).\footnote{Our mechanisms can also be extended to other codes similarly. MDS codes can make  the expected overall runtime ${\Theta}(\log n)$ times	faster than  uncoded matrix multiplication (e.g.,  \cite{lee2017speeding}).} Then, the load assignment and the corresponding expected overall runtime are given as follows:

\begin{lemma}
	\label{lm2}
	Under the $(n, k)$-MDS codes,   each worker has the same computation load 
\begin{equation}
\label{14}
\ell=\frac{r}{k}.
\end{equation}
The expected overall runtime under \eqref{14}  is 
\begin{equation}
\label{time}
\begin{split}
\mathbb{E}[T ]=\frac{r}{k}\hspace{-0.5mm}\left(a\hspace{-0.5mm}+\hspace{-0.5mm}\frac{H_{n}\hspace{-0.5mm}-\hspace{-0.5mm}H_{n-k}}{\mu}\right)  \simeq\frac{r}{k}\hspace{-0.5mm}\left(a\hspace{-0.5mm}+\hspace{-0.5mm}\frac{1}{\mu}\log\frac{n}{n\hspace{-0.5mm}-\hspace{-0.5mm}k}\right),
\end{split}
\end{equation}
where $H_{n} \triangleq \sum_{i=1}^{n} \frac{1}{i} $, and $H_{n}  \simeq \log n$ when $n$ is large.
\end{lemma}
 Proof of Lemma \ref{lm2} is given in Appendix V of the technical report \cite{appen1}. 
The analysis of workers' participation decisions and the platform's  strategies is similar to that in Section \ref{hetero}. Then, the platform's optimal targeted worker type set, optimal rewards, and optimal recovery threshold in the ``Incomplete-HeteCostOnly'' scenario are given in Proposition \ref{thmb}:


\begin{proposition}
	\label{thmb}
	Given $ {c_1} \le ...\le    {c_{M}} $ in the ``Incomplete-HeteCostOnly'' scenario,  there exists a threshold type $n^{IncHco}$,
		\begin{equation}
	\label{19}
	n^{IncHco}  =\arg\min_{{n\in \mathcal{M}}} \frac{\gamma_1+\gamma_2{c_n}\sum_{m=1 }^n N_m}{\sum_{m=1 }^n N_m},
	\end{equation}
	such that it is optimal for the platform to induce the participation of worker types in the following targeted set $\mathcal{S}^{IncHco}  $:
			\begin{equation}
	\begin{split}
	\mathcal{S}^{IncHco}  =\{n|n\in \mathcal{M},n\leq n^{IncHco}\}.
	\end{split}
	\end{equation} 
	The optimal expected reward  for a type-$m\in \mathcal{M}$ worker is
	\begin{equation}
	p_m^{IncHco}={c}_{n^{IncHco}  }\frac{r}{k^*}\left(a+\frac{1}{\mu}\log\frac{1}{1-\alpha}\right),\forall m\in \mathcal{M}, 
	\end{equation}
	and the platform's optimal recovery threshold is 
	\begin{equation}
	k^*=\alpha N^{(\mathcal{S}^{IncHco}  )},
	\end{equation}
	where  $\alpha\triangleq \left[1+\frac{1}{W_{-1}\left(-e^{-a\mu-1}\right)}\right] \in [0,1]$,   $W_{-1}(\cdot)$ is the lower branch of Lambert W function\footnote{$x=-W_{-1}\left(-\frac{1}{t}\right)$ is the unique solution to $\frac{e^{x}}{x}=t, t \geq e$ and $x \geq 1$.}, and $N^{(\mathcal{S}^{IncHco})}\triangleq\sum_{m \in \mathcal{S}^{IncHco}}N_m$. 
\end{proposition}

\begin{figure}[tbp]
	\centering
	\includegraphics[width=0.5\linewidth]{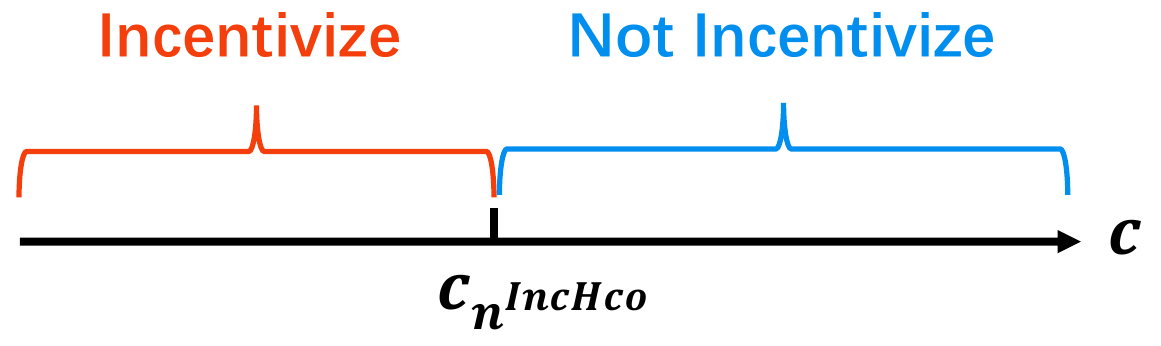}
	\vspace{-2.5mm}
	\caption{Illustration of $\mathcal{S}^{IncHco}$ in ``Incomplete-HeteCostOnly'' scenario.}
	\label{figa}
	\vspace{-2.5mm}
\end{figure}
Proof of Proposition \ref{thmb} is given in Appendix VI of the technical report \cite{appen1}. 

As illustrated in Fig.~\ref{figa}, Proposition \ref{thmb} shows that  in the ``Incomplete-HeteCostOnly'' scenario, the platform   prefers  the worker types with small marginal costs, as workers have the same computation performance in this case. 
The platform gives all workers the same expected reward, which ensures that worker types in the desirable set $\mathcal{S}^{IncHco}$ will participate, and 
types not  in $\mathcal{S}^{IncHco}$ will  not participate. 

More importantly, Proposition \ref{thmb}   presents that the MDS codes' optimal recovery threshold  $k^*$ is linearly proportional to the total participator number $N^{(\mathcal{S}^{IncHco}  )}$, which provides an easy-to-implement guideline for data encoding. According to \eqref{14}, the optimal load assignment is $\ell_{m}^*=r/k^*, \forall m\in \mathcal{S}$.

Next, we will compare the platform's strategies and costs in different scenarios through numerical analysis in Section \ref{simulation}.
 

\section{Simulation Results}
\label{simulation}
In Section \ref{vs}, we  will   evaluate the performance of our proposed incentive mechanisms. 
In Section \ref{more},  we will further consider  a \emph{strongly incomplete information scenario} where the platform further lacks the information about workers' distributions in different types (i.e., $N_m$ of  type $m$). 
We will  show that our incentive mechanism for incomplete information scenario is asymptotically optimal under strongly incomplete information.
 


\begin{table}[tbp]
	\caption{Workers' Parameter Settings in Simulation}  
	\vspace{-4.5mm}
	\begin{center}
		\begin{tabular}{|c|c|}
			\hline
			$(c_1,\mu_1,a_1)=(1,50,0.012)$ &  $(c_2,\mu_2,a_2)=(7,100,0.024)$ \\
			\hline
			$(c_3,\mu_3,a_3)=(8,200,0.033)$ & $(c_4,\mu_4,a_4)=(3,10,0.031)$ \\
			\hline
			$(c_5,\mu_5,a_5)=(16,400,0.040)$ &  $(c_6,\mu_6,a_6)=(5,20,0.081)$  \\
			\hline
			$(c_7,\mu_7,a_7)=(21,800,0.044)$ & $(c_8,\mu_8,a_8)=(9,40,0.123)$    \\
			\hline
			$(c_9,\mu_9,a_9)=(12,80,0.153)$ &  $(c_{10},\mu_{10},a_{10})=(20,160,0.172)$  \\
			\hline
		\end{tabular}
		\label{3tab} 
		\vspace{-6mm}
	\end{center}
\end{table}
When workers have three-dimensional heterogeneity in computation performances and costs, we consider  $M= 10$ types of workers with parameter settings in Table \ref{3tab},    
   similar to that in \cite{reisizadeh2019coded}.  
 The parameters  satisfy increasing  cost-performance ratio relationship:   $\Omega_1\le ...\le   \Omega_M$. 
 There are $r=1000$ inner products to be finished by workers. 
  Each  type has $N_m=N/M,\forall m\in \mathcal{M}$ workers.   We consider both $\gamma_1=2000$ and $\gamma_2=1$ to reflect the platform's different evaluation scenarios  on time cost and payment.

\subsection{Complete versus Incomplete Information}
\label{vs}
\begin{figure}[tbp]
	\centering
	\includegraphics[width=2.2  in]{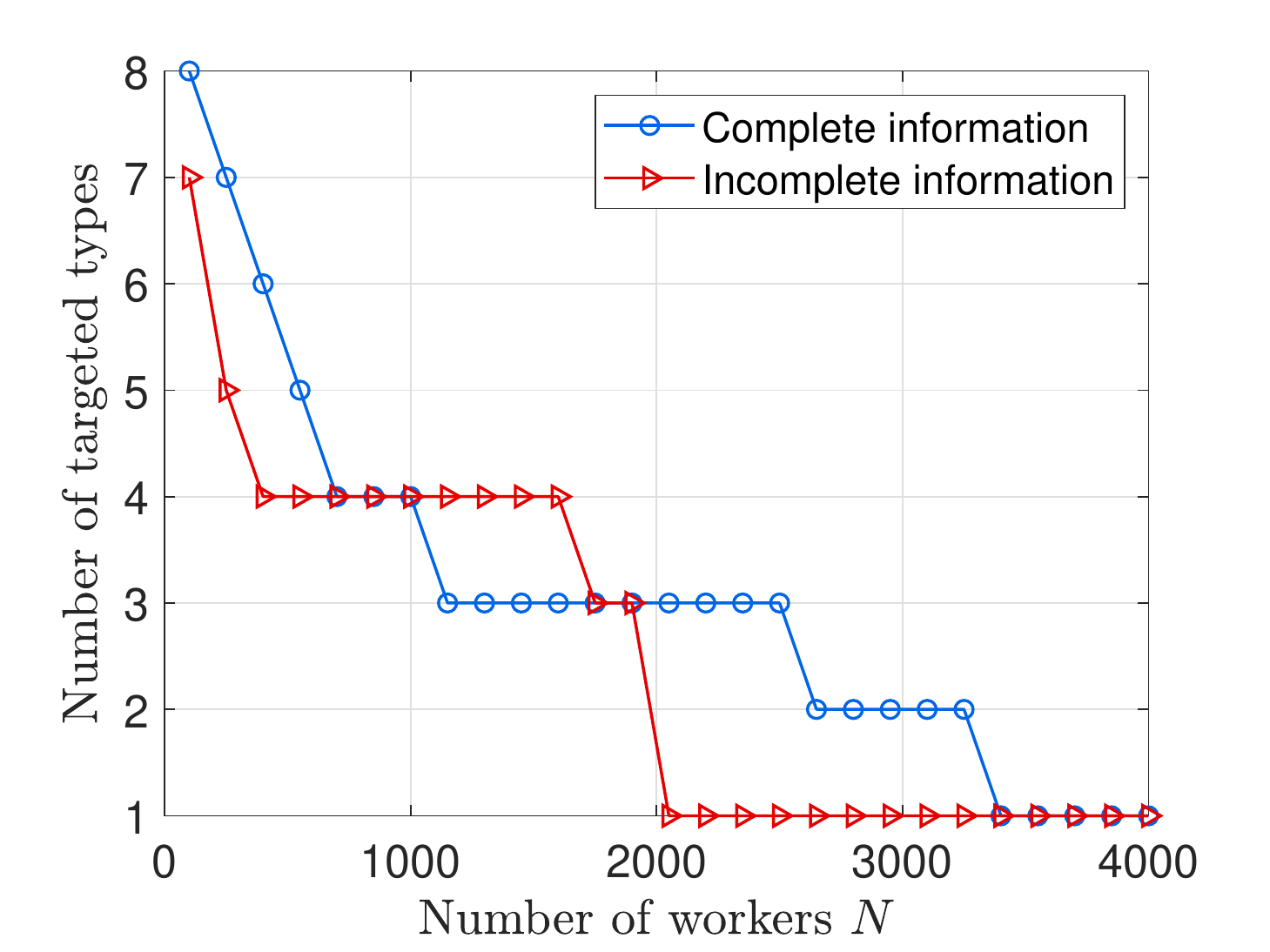}
	\vspace{-3mm}
	\caption{Number of worker types targeted by the platform (under both complete and incomplete information)  versus the total worker number $N$.}
	\label{fig10}
	\vspace{-3.5mm}
\end{figure}

Fig.~\ref{fig10} shows the   number of platform's targeted worker types   characterized  in Theorems \ref{thmc} and \ref{thmd} versus the total worker number $N$, under both complete and incomplete information. Both curves are decreasing in $N$. This is because  the    worker types of small cost-performance ratios have more workers when $N$ increases.  The platform can rely less on the  inefficient or costly types. 
By comparing the results in two information scenarios in Fig.~\ref{fig10}, we realize that the platform does not always target at a smaller group of workers under incomplete information. For example, the platform includes $4$ worker types under incomplete information versus $3$ worker types under complete information at around $N=1400$. The platform under incomplete information afraid of large overall runtime wants to include one more type   despite the mildly increased cost.

\begin{figure}[tbp]
	\centering
	\includegraphics[width=2.2  in]{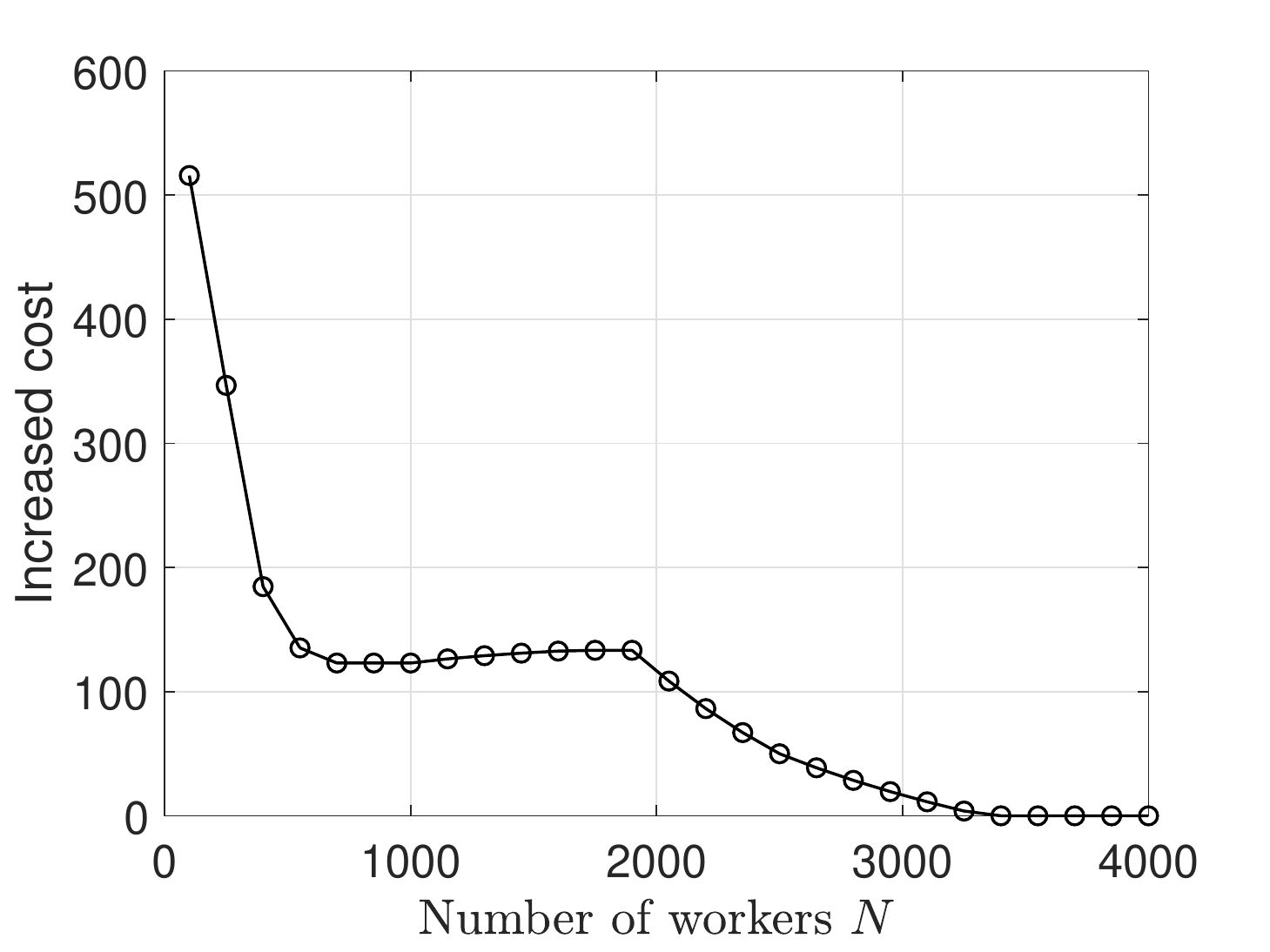}
	\vspace{-3mm}
	\caption{Platform's   increased cost  due to the lack of information versus the total worker number $N$.}
	\label{fig8}
	\vspace{-1mm}
\end{figure}
\begin{figure}[tbp]
	\centering
	\vspace{-2mm}
	\includegraphics[width=2.2  in]{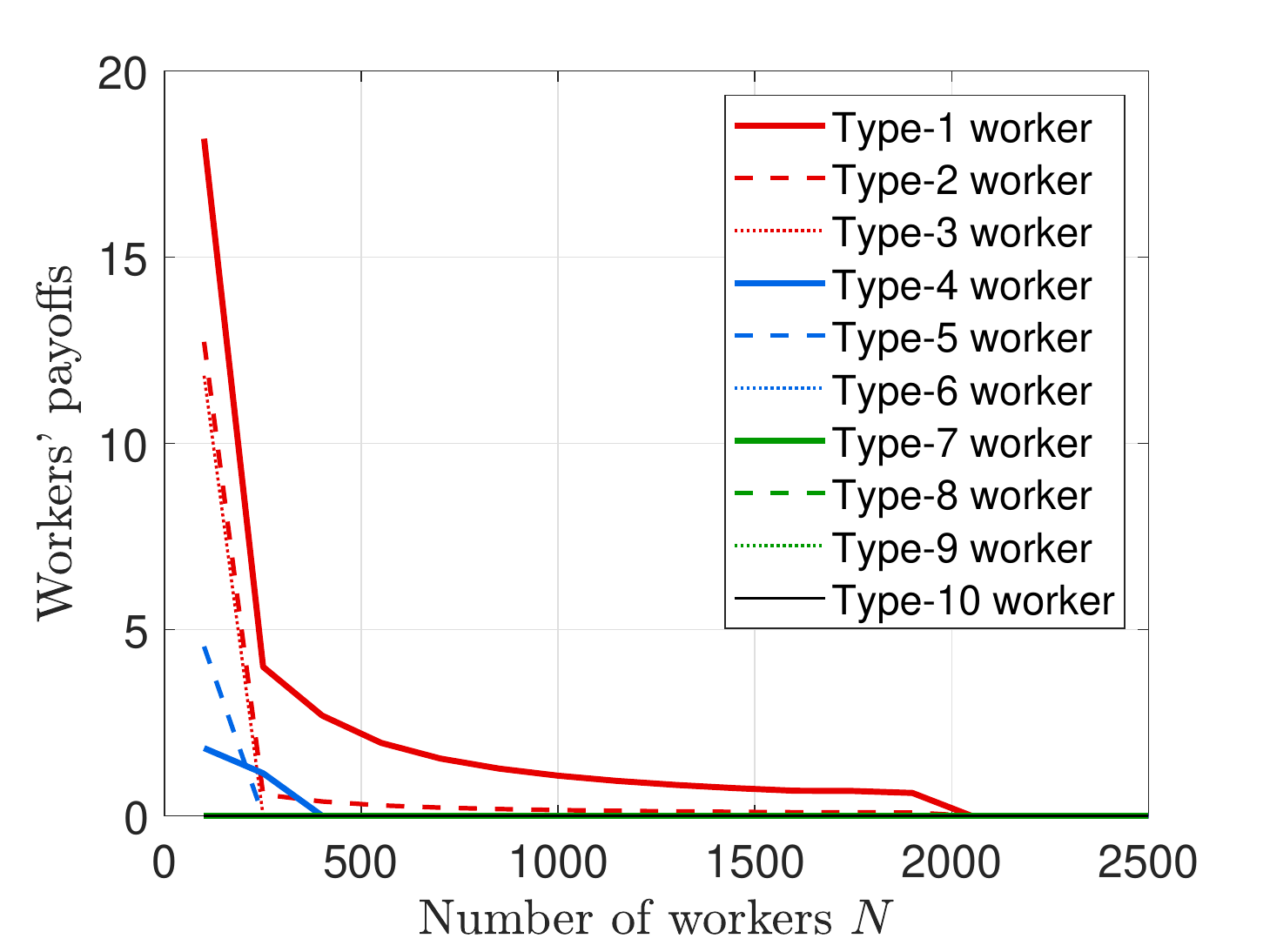}
	\vspace{-3mm}
	\caption{Different worker types'  payoffs under incomplete information versus the total worker number $N$.}
	\label{fig13}
	\vspace{-5mm}
\end{figure}
\begin{figure}[tbp]
	\centering
	\includegraphics[width=2.2  in]{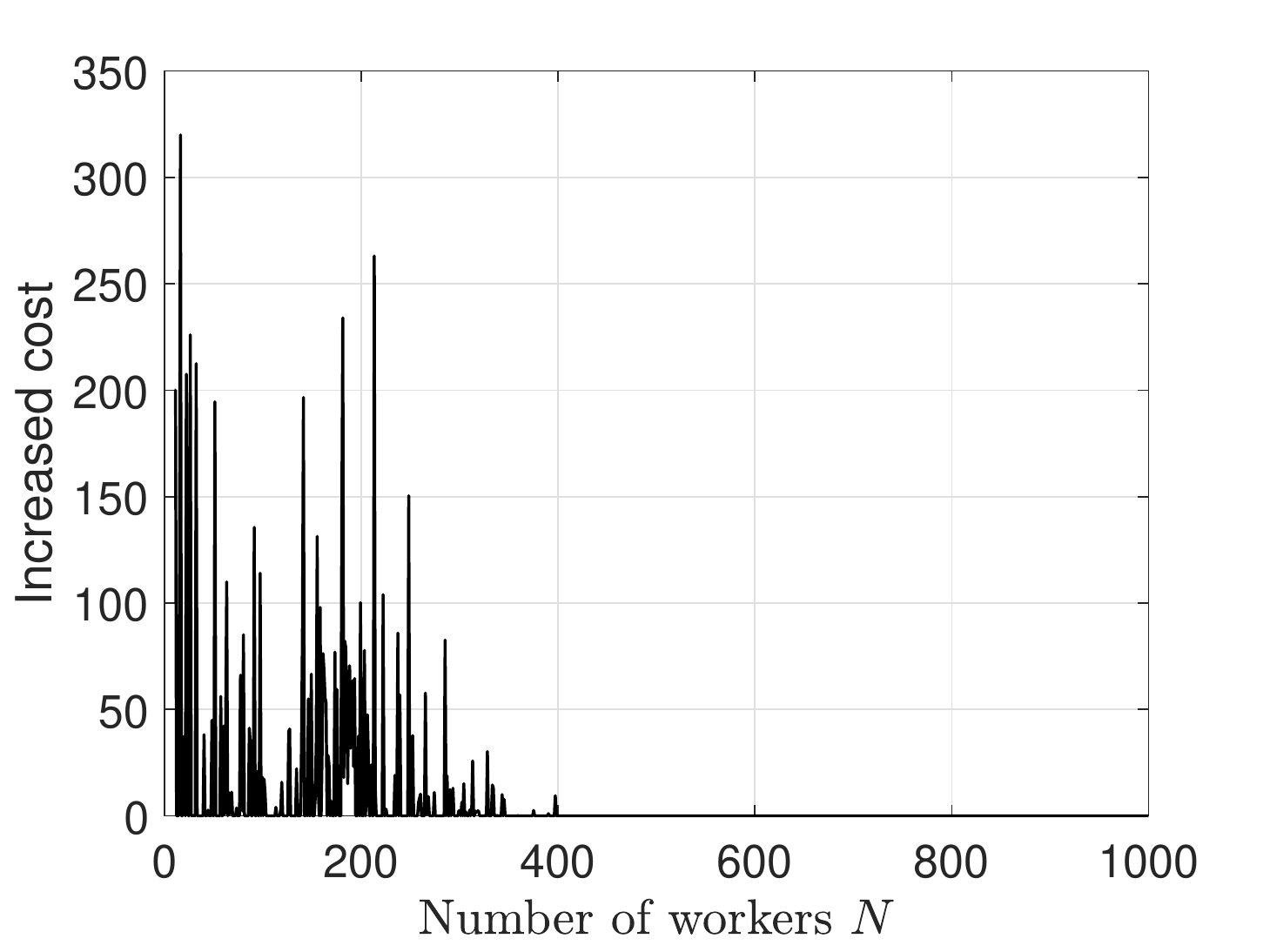}
	\vspace{-3mm}
	\caption{Platform's  increased cost due to strongly incomplete information compared with incomplete information versus the total worker number $N$.}
	\label{fig15}
	\vspace{-5mm}
\end{figure}

Fig.~\ref{fig8} compares the platform's cost objectives under complete and incomplete information,  by measuring increased cost due to incomplete information  versus the worker number $N$.   When the number of workers becomes very large  (larger than $3400$ in Fig.~\ref{fig8}), the platform   obtains the same cost in these two scenarios. 
This is because  the platform only chooses the type with smallest cost-performance ratio in both complete and incomplete information scenarios (Fig.~\ref{fig10}). 
However, the platform's  cost gap does not monotonically decrease in $N$. The platform may sometimes target    more worker types under incomplete information (see Fig.~\ref{fig10} when $N$ equals $1400$), which   increases cost difference   compared to the complete information scenario.

Fig.~\ref{fig13} further shows that different types of workers' payoffs under incomplete information versus the worker number $N$. Only the targeted worker types can have positive payoffs, and such payoffs are overall decreasing in $N$ due to mutual competition and the platform's smaller set of targeted   types. 


%

\subsection{Extension of Incomplete Information}
\label{more}
In Section \ref{incomple}, we have studied the mechanism design under incomplete information without knowing each worker's computation time and cost. 
Here we further evaluate the performance of the proposed mechanism  in Theorem \ref{thmd} under strongly incomplete information, where the platform only knows the total number of workers and the distribution of each worker's type. Each worker has an equal probability of 10\% of belonging to each worker type. 

In this strongly incomplete information scenario, as  the number of workers $N$ becomes  large, according to the law of large numbers,  the empirical  number of each type of workers approaches to the expected value computed based on the distribution. 
 As shown in Fig.~\ref{fig15}, when $N$ is larger than $400$, the increased cost due to strongly incomplete information reaches zero\footnote{The fluctuations are due to the random realization of the number of each type workers.}. Thus, our incentive mechanism in Theorem \ref{thmd} performs optimally under this strongly incomplete information scenario as $N$ goes to infinity. 



\section{Conclusion}
\label{conclusion}
This paper  studied the important incentive issues  in coded machine learning. We  captured the trade-off between the platform's time cost and payment  in his payoff function. We   proposed the optimal incentive mechanisms for the workers in both complete information and incomplete information scenarios, under workers' multi-dimensional heterogeneity in computation performances and costs. 
In the presence of the high complexity of the worker selection problem, we managed to summarize  workers' multi-dimensional heterogeneity into a one-dimensional metric and solved the   problem in a linear complexity. We also demonstrated that the optimal recovery threshold is linearly proportional to the total participator number when using MDS codes. We showed that compared with the complete information scenario, the effect of incomplete information on the platform's cost will disappear when the total worker number is sufficiently large.
We will further study the optimal incentive mechanisms of competing platforms in future work.

\bibliographystyle{IEEEtran}
\bibliography{ref} 

\end{document}